\documentclass[prb,showpacs,twocolumn,superscriptaddress,aps,a4paper]{revtex4}
\usepackage{float}
\usepackage{dcolumn}
\usepackage{amsmath}
\usepackage{graphicx}
\usepackage{latexsym}
\usepackage{amsfonts}
\usepackage{amssymb}
\usepackage{bbm}
\usepackage{pst-node}

\DeclareGraphicsExtensions{.pdf,.gif,.jpg}

\def\a{\alpha}
\def\b{\beta}

\def\g{\gamma}

\def\s{\sigma}

\def\w{\omega}

\def\ve{\varepsilon}
\def\D{\Delta}
\def\calE{{\cal E}}

\def\ket{\rangle}

\def\ra{\rightarrow}
\def\de{\partial}
\def\inf{\infty}
\def\ua{\uparrow}
\def\da{\downarrow}

\newcommand{\be}{\begin{equation}}
\newcommand{\ee}{\end{equation}}
\newcommand{\beq}{\begin{eqnarray}}
\newcommand{\eeq}{\end{eqnarray}}

\begin{document}

\title{Missing derivative discontinuity of the exchange-correlation 
energy for attractive interactions: the charge Kondo effect}

 \author{E. Perfetto}
 \affiliation{Dipartimento di Fisica, Universit\`a di Roma 
 Tor Vergata, Via della Ricerca Scientifica 1, I-00133 Rome, Italy}

 \author{G. Stefanucci}
 \affiliation{Dipartimento di Fisica, Universit\`a di Roma
 Tor Vergata, Via della Ricerca Scientifica 1, I-00133 Rome, Italy}
 \affiliation{INFN, Laboratori Nazionali di Frascati, Via E. Fermi 40,
 00044 Frascati, Italy}
 \affiliation{European Theoretical Spectroscopy Facility (ETSF)}

\begin{abstract}

We show that the energy functional of ensemble Density Functional 
Theory (DFT) [Perdew et al., Phys. Rev. Lett. {\bf 49}, 1691 (1982)] in 
systems with attractive interactions is a convex 
function of the fractional particle number $N$ and is given by a series of 
straight lines joining a {\em subset} of ground-state energies. 
As a consequence the exchange-correlation (XC) potential is not discontinuous 
for all $N$. We highlight the importance of this exact result
in the ensemble-DFT description of the negative-$U$ Anderson model. 
In the atomic limit the discontinuity of the XC potential
is missing for odd $N$ while for finite hybridizations the 
discontinuity at even $N$ is broadened. 
We  demonstrate that the inclusion of these properties in any 
approximate XC potential is crucial to reproduce the 
characteristic signatures of the charge-Kondo effect in the conductance and charge 
susceptibility.


\end{abstract}


\maketitle

\begin{figure}[t]
\includegraphics[width=7.cm]{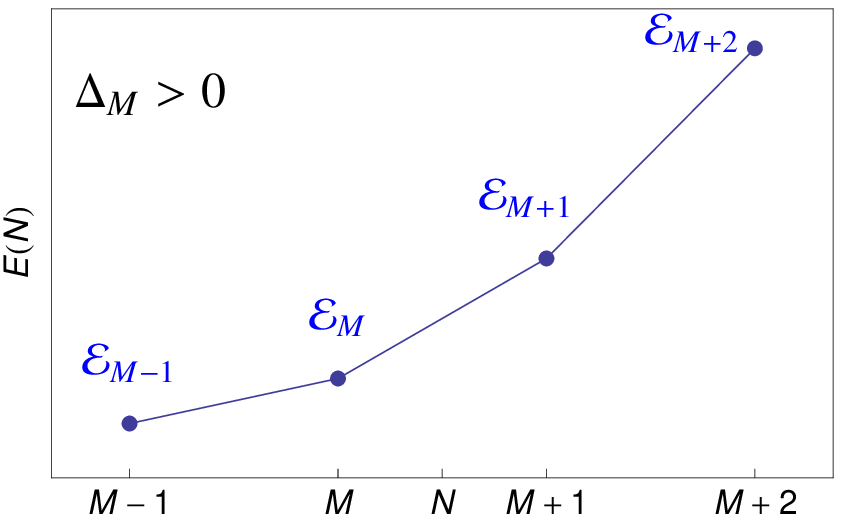}
\includegraphics[width=7.cm]{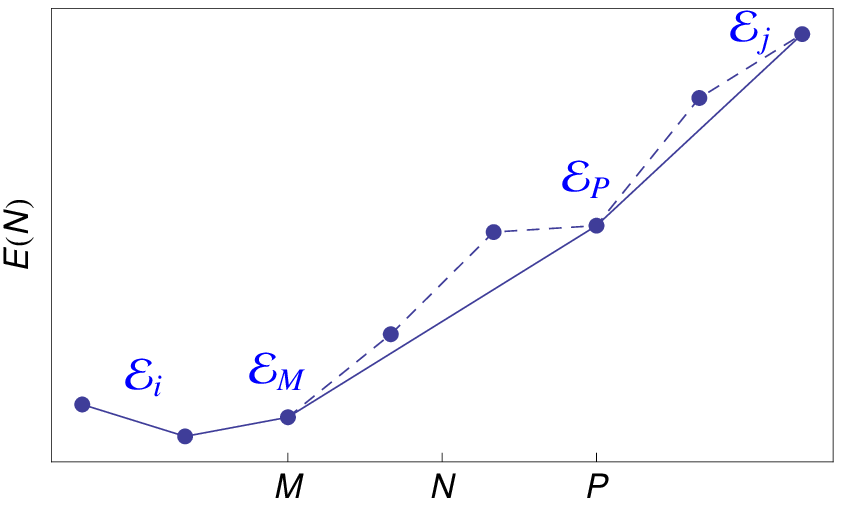}
\caption{Illustrative examples of the total energy $E(N)$ (solid line) as a function 
of the fractional number $N$ when the convexity inequality (\ref{convex}) 
is  (top panel) and is not (bottom panel) fulfilled. 
In the bottom panel the dashed line 
joins consecutive ground-state energies and differs from $E(N)$.}
\label{fig1}
\end{figure}

Density functional theory\cite{hk.1964,ks.1965} (DFT) provides a rigorous and computationally viable
tool to calculate the electronic properties of many-particle interacting systems. 
In spite of the great success in a wide range of applications, its 
practical use is still problematic  in systems with fluctuating number of particles.
Popular approximations like LDA or GGA are inadequate to predict, e.g., the band gap of 
solids,\cite{levy,sham} the 
correct dissociation of heteroatomic molecules\cite{perdew1,perdewchp,hrg.2012} or the electrical conductivity
of  nanoscale junctions.\cite{se.2008,mera} A conceptual advance to 
deal with these cases is the ensemble-DFT put forward by Perdew et 
al..\cite{perdew1} These authors  extended the 
original DFT formulation\cite{hk.1964,ks.1965} to a fractional number 
$N$ of electrons, and pointed out the non-differentiability
of the energy functional 
$E[n]$ of the density $n$ at integers $N=\int n$. Typically the discontinuity in $\de E/\de N$ 
is the difference between the ionization energy and the 
electron affinity since for any $N$ between two consecutive 
integers $M$ and $M+1$ one has 
\be
E(N)= (M+1-N) \calE_{M} + (N-M)\calE_{M+1},
\label{perdew}
\ee
i.e., $E(N)$ is a series of straight lines  joining {\em consecutive} 
ground-state energies $\calE_{M}$ of the isolated
system  with $M$ particles. Fig. \ref{fig1} (top panel) illustrates 
the typical outcome of a ground state calculation of $E(N)$.
It is worth recalling that the crucial hypothesis for the validity of 
Eq. (\ref{perdew}) is the convexity inequality   
\be
\D_{M} \equiv \calE_{M+1} + \calE_{M-1} -2\calE_{M} \geq 0.
\label{convex}
\ee
Indeed, in this case one can show that
the density matrix which minimizes the total energy is a 
linear combination of projection operators over the ground states 
with $M$ and $M+1$ particles.
As discussed in Ref. \onlinecite{perdew1} the hypothesis $\D_{M}\geq 0$
is certainly reasonable in systems with repulsive interactions.
In contrast the convexity inequality can be violated 
in the attractive case. For instance in the attractive Hubbard 
model $\D_{M}$ is positive for even $M$ and negative 
otherwise.\cite{hu,kulik}
What consequences does the 
break-down of Eq. (\ref{convex}) have in ensemble-DFT? What are the 
physical implications?

In this paper we generalize Eq. (\ref{perdew}) to arbitrary $\D_{M}$.
In particular we prove that $E(N)$ is a
convex function given by a 
series of straight lines joining 
a {\em subset} of ground-state energies, as schematically illustrated 
in Fig. \ref{fig1} (bottom panel). 
We further study the implications of the 
missing derivative discontinuity in the negative-$U$ Anderson model.
In this prototype system the attractive interaction
is at the origin of the so called charge-Kondo effect:\cite{haldane,coleman}
at very low temperature the fluctuations 
between the empty state and the doubly-occupied state of the impurity 
level
produce a strong enhancement of the charge susceptibility, 
accompanied by a drastic shrinkage of the conductance peak.
We show that the key features of the 
charge-Kondo effect can be captured within ensemble-DFT. We extend a 
recently proposed functional\cite{stef} devised for the spin-Kondo effect to attractive interactions 
and account for the broadening of the discontinuity\cite{evers} due to the 
finite hybridization of the impurity level.
The transition from the spin-Kondo effect  
to the charge-Kondo effect is caused by the shift of the
discontinuity of the exchange-correlation (XC) potential from $N=1$ at 
$U>0$\cite{lsoc.2003,kurth} to 
$N=0$ and $N=2$ at $U<0$.

\underline{\em Theorem}:
Given the ground-state 
energies $\{\calE_{I}\}$
of the isolated system with $I$ particles, if
\be
\frac{\calE_{I}-\calE_{M}}{I-M}<\frac{\calE_{P}-\calE_{M}}{P-M}<\frac{\calE_{J}-
\calE_{M}}{J-M}
\label{hyp}
\ee
for every $I<M$ and every $J>M$, then in the range $M<N<P$ it holds
\be
E(N)=\frac{P-N}{P-M}\calE_{M}+\frac{N-M}{P-M}\calE_{P}.
\label{general}
\ee
Graphically this means that for  $N 
\in [M,P]$ the energy $E(N)$ lies on the straight 
line connecting $\calE_{M}$ to $\calE_{P}$ {\it if and only if} the slope 
$\frac{\calE_{P}-\calE_{M}}{P-M}$ is larger than all the slopes of the lines
connecting $\calE_{M}$ to $\calE_{I<M}$ and smaller
than all the slopes of the lines connecting $\calE_{M}$ to 
$\calE_{J>M}$, see Fig. \ref{fig1} bottom panel.

\underline{\em Proof}: We have to show that in the range  $M<N<P$ the
variational energy $E_{var}(N)=\sum_{L}\w_{L}\calE_{L}$ cannot be 
smaller than
the energy $E(N)$ in Eq. (\ref{general})  for any $\{\w_{L}\}$ 
constrained to satisfy 
\be
\sum_{L}\w_{L}L= N, \quad\quad
\sum_{L}\w_{L} = 1, 
\ee
and $\w_{L}\geq 0$ for all $L$.
Using Eq. (\ref{hyp}) one has
\beq
E_{var}(N)&>&\w_{M}\calE_{M}
\nonumber \\
&+&
\sum_{I<M}\w_{I}\left[\calE_{M}+(\calE_{P}-\calE_{M})\frac{I-M}{P-M}\right]  \nonumber 
\\
&+& 
\sum_{J>M}\w_{J}\left[\calE_{M}+(\calE_{P}-\calE_{M})\frac{J-M}{P-M}\right] \nonumber\\
&=& \calE_{M}+ 
\frac{\calE_{P}-\calE_{M}}{P-M}\sum_{L}\w_{L}(L-M)=E(N), \quad\;\;
\eeq
which proves the theorem.

\begin{figure}[t]
\includegraphics[width=7.2cm]{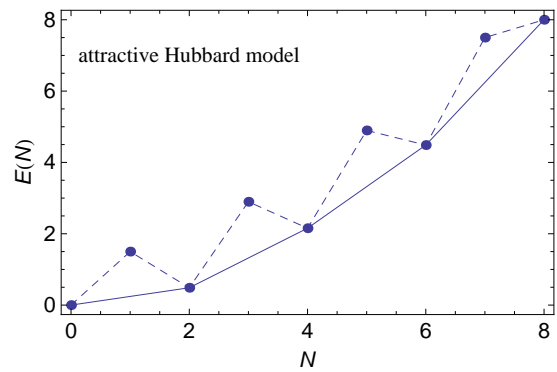}
\caption{$E(N)$ of Eq. (\ref{general}) (solid line)
for the 4-sites attractive Hubbard ring with
$U=-5$ and on-site energy $\ve_{d}=3.5$. 
Energies are in 
units of the hopping integral. The dashed line 
joins consecutive ground-state energies.}
\label{fig2}
\end{figure}

Thus $E(N)$ in Eq. (\ref{general}) is a convex function of $N$ and reduces
to Eq. (\ref{perdew}) provided that the convexity inequality 
$\D_{M}\geq 0$ is satisfied
for all $M$ since in this case  $P=M+1$. The physical content of the 
theorem is clear. If the system is open to  a charge reservoir the 
density matrix at zero temperature is a mixture of ground states with 
$M$ and $P$ particles.
For example in Fig. \ref{fig2} we show that
in the attractive Hubbard ring $P=M+2$, see also Refs. 
\onlinecite{hu,kulik}. The value 
$P=M+2$ is peculiar of attractive systems where the electron pairing 
causes $\Delta_{M}\lessgtr 0$ for even/odd $M$. 
This property is consistent with the experimental observation of the
Coulomb blockade of Cooper
pairs\cite{devoret,gunn} (Cooper staircase) 
in  superconducting single-electron transistors, where a superconductive 
island is connected to metallic leads. In this situation the application 
of a gate voltage $v_{g}$ to the 
attractive region causes a jump of 2 in the  number of particles
at the special values 
$v_{g}=\frac{\calE_{2M}-\calE_{2M+2}}{2}$.\cite{perfetto}

In ensemble-DFT the 
discontinuity $\de E/\de N$ is the sum the Kohn-Sham (KS) discontinuity, which is 
zero for odd $N$, and the XC discontinuity $\D_{\rm xc}(N)$. Since 
$\de E/\de N=0$ for odd $N$ we conclude that
\be
\D_{\rm xc}(N)=0 \quad\quad \textrm{for odd $N$}.
\ee

In the following, we consider a negative-$U$ Anderson model as
an example in which XC discontinuity is missing. The Hamiltonian
describes a set of noninteracting electrons coupled to a site at which
Hubbard type interaction occurs.\cite{haldane,coleman}  This is 
an effective model for conduction 
electrons coupled to an interacting impurity with vibrational modes.
For strong  electron-phonon coupling
the polaronic shift can overcome the Coulomb charging energy and the 
effective electron-electron interaction turns out to be attractive.
The Hamiltonian reads, in standard notation,
\beq
H&=&t\sum_{\a, \s} \sum_{i=1}^{\infty}  [c^{\dagger}_{i\a\s}c_{i+1\a\s} 
+h.c.] + v_{g}\sum_{\s}n_{d \s} \nonumber \\
&+& Un_{d\ua}n_{d\da} + t'\sum_{\a,\s}[c^{\dagger}_{1\a\s}d_{\s}+h.c.],
\label{hand}
\eeq
where  $t$ is the nearest-neighbour hopping in the leads, $t'$ is the 
lead-impurity hopping, $U<0$ is the attractive interaction, and $v_{g}$ 
is the gate voltage coupled to the impurity density $n_{d\s}=d_{\s}^{\dag}d_{\s}$.
Below we focus on the half-filled system and hence take the chemical 
potential $\mu=0$. At very low temperature and  gate 
voltage around $\bar{v}_{g}=-U/2=|U|/2$ this model exhibits the so 
called charge-Kondo effect.\cite{haldane,coleman}
This effect consists in the formation of a 
``local pair'' at the impurity and is due to  strong 
charge fluctuations between the nearly degenerate states $|0\ket$ and 
$|\ua\da\ket$ of the empty and doubly occupied $d$-level.
As predicted by Taraphder and Coleman\cite{coleman} the 
local pair  is ``screened'' by the surrounding conduction electrons 
and forms  an ``isospin singlet''. With  increasing $|U|$  the
main features of
the charge-Kondo effect  are i)
the shrinkage of the conductance resonance at $v_{g}=\bar{v}_{g}$
and ii) the large growth of the  charge 
susceptibility $\chi_{d}=-\partial n_{d}/\partial v_{g}$.\cite{coleman,mravlje}
These results can be qualitatively understood by mapping the 
Hamiltonian of Eq. (\ref{hand}) into the positive-$U$ Anderson
model. Under a particle-hole transformation in the spin 
down sector, $d_{\da}\ra d^{\dag}_{\da}$ and  
$c_{i\a\da}\ra (-1)^{i}c^{\dag}_{i \a \da}$, the original Hamiltonian is 
transformed into the  positive-$U$ Anderson Hamiltonian
with fixed gate voltage $-|U|/2$ and  effective 
magnetic field $B_{\mathrm{eff}}=-|U|/2+v_{g}$ coupled to 
$(n_{d\ua}-n_{d\da})$.\cite{vonoppen1,vonoppen2,lopez}
Since the magnetic field suppresses very 
efficiently the Kondo 
correlations\cite{wiegmann,vonoppen1,vonoppen2,mravlje,arrachea,cornaglia1,cornaglia2} 
the spin-Kondo 
effect in the transformed Hamiltonian occurs only in the
proximity of $v_{g}=\bar{v}_{g}$. Consequently the conductance drops rapidly to zero
as $v_{g}$ deviates from $\bar{v}_{g}$. At resonance 
the spin fluctuations in the transformed Hamiltonian 
correspond to  ``isospin'', i.e., charge,
fluctuations in the original Hamiltonian, thus leading to 
the formation of an isospin singlet (local pair). This phenomenology 
explains the large growth of the charge susceptibility 
$\chi_{d}$ as $|U|$ increases (this growth
is not observed for positive $U$).  

\begin{figure}[t]
\includegraphics[width=7.2cm]{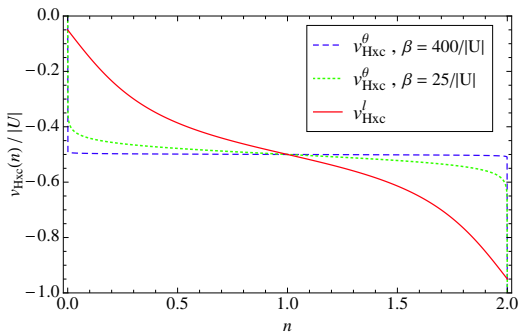}
\caption{Potential $v^{\theta}_{\rm Hxc}$ at very low temperature, 
$\b=400/|U|$, and comparison between $v^{\theta}_{\rm Hxc}$ at 
$\b=1/W\gamma \approx 25/|U| $ and $v^{l}_{\rm Hxc}$ at zero temperature with 
$\g=0.125|U|$. The Lorentzian 
broadening is much larger than the thermal broadening for $\g\ll |U|$. }
\label{vxc}
\end{figure}

Let us show how these features can be captured in ensemble-DFT.
In a recent Letter\cite{stef} an approximate Hartree-XC potential for the 
positive-$U$ Anderson model was proposed. The exact energy functional of the 
isolated impurity reads
\be
v^{\theta}_{\rm Hxc}(n_{d})=\frac{U}{2}+g(n_{d}-1),
\label{hxcpot}
\ee
where
\be
g(x)=\frac{U}{2}+\frac{1}{\b}\log\frac{x+\sqrt{x^{2}+e^{-\b 
U}(1-x^{2})}}{1+x},
\ee
$\beta$ being the inverse temperature. 
For $U>0$ and in the limit $\b \to \infty$ the potential 
$v^{\theta}_{\rm Hxc}(n_{d})\ra U\theta(n_{d}-1)$ which has a discontinuity $U$ 
at $n_{d}=1$.
In the Wide Band 
Limit Approximation (WBLA) $t,t'\to\inf$ with constant $2t'^{2}/t=\gamma\ll 
U$ (weak tunneling rate) one can approximate the Hartree-XC potential on the 
impurity $v_{\rm Hxc}\approx v_{\rm Hxc}^{\theta}$
and set it to zero in the leads.\cite{stef}
The discontinuity forces the occupation 
to be unity for gate voltages $0<v_{g}<U$.\cite{kurth,tfsb.2005}  
Thus the KS potential is pinned
at the Fermi energy and the KS conductance exhibits a 
Kondo plateau as a function of $v_{g}$.\cite{stef,burke,evers2}
\begin{figure}[t]
\includegraphics[width=7.2cm]{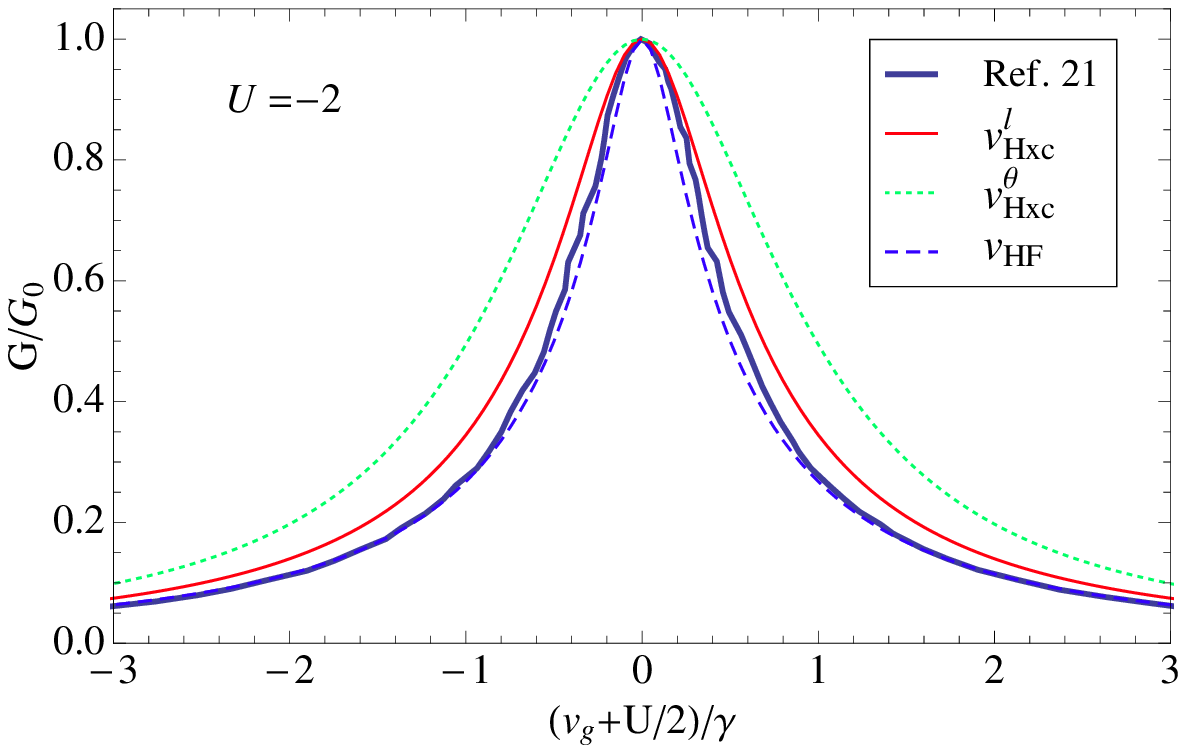}
\includegraphics[width=7.2cm]{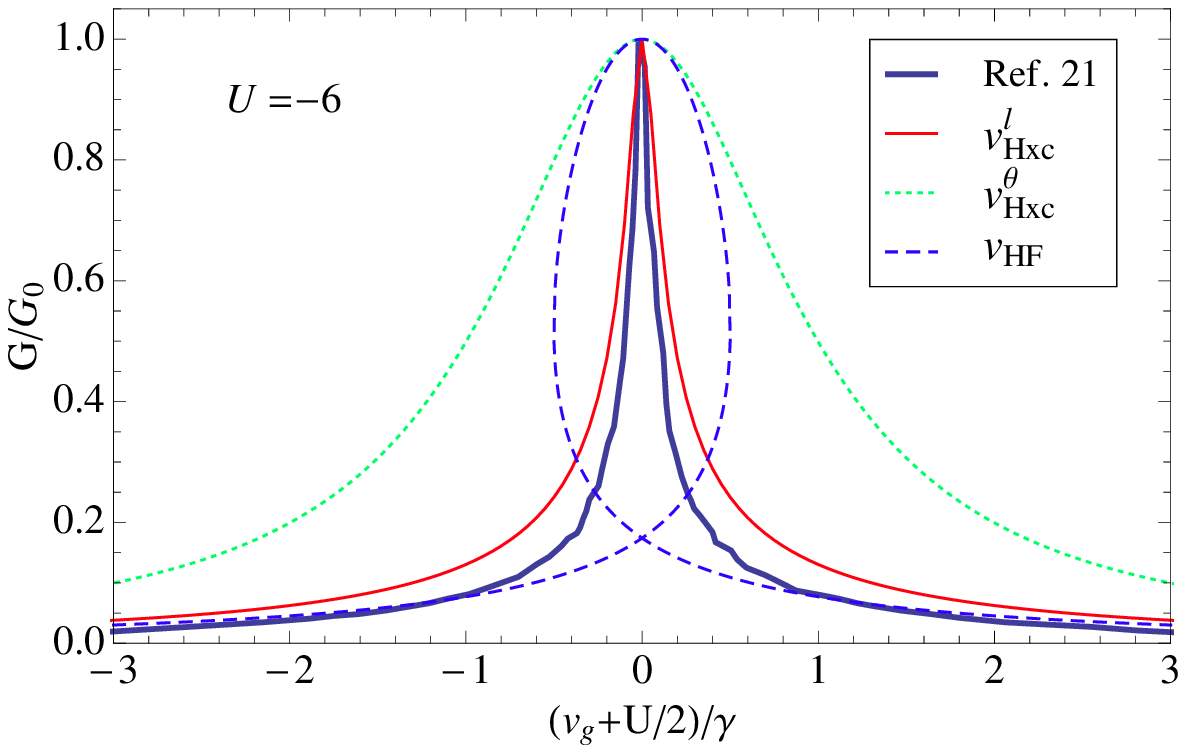}
\caption{Conductance $G$ as a function of $v_{g}$
for two values of $U=-2,-6$ (in units of $\gamma$) and different approximation schemes.
The data is compared to the variational results
of Ref.  \onlinecite{mravlje}, which agree closely with the NRG data 
of Ref. \onlinecite{cornaglia1}, and can
be, therefore, considered as exact.}
\label{comparison}
\end{figure}

The physical argument leading to Eq. (\ref{hxcpot}) is independent of the sign 
of $U$ and we may argue that the functional $v_{\rm 
Hxc}^{\theta}$ should predict, at least qualitatively, the correct 
conductance also for negative $U$. In the analysis below we consider 
the zero-temperature case.
For $U<0$ the potential  $v_{\rm 
Hxc}^{\theta}$ is not discontinuous at $n_{d}=1$ but instead develops
two discontinuities (of size $|U|/2$)
at $n_{d}=0$ and $n_{d}=2$, see Fig. \ref{vxc}.\cite{capelle}
Within the WBLA we determine the occupancy on the 
impurity by solving the self-consistent equation 
$n_{d}=1-\frac{2}{\pi}\tan^{-1}\!\left[\frac{v_{g}+v_{\rm 
Hxc}(n_{d})}{\g}\right]$ with $v_{\rm Hxc}=v^{\theta}_{\rm Hxc}$. Once $n_{d}$ is known we calculate the 
KS conductance from
\be
\frac{G}{G_{0}}=\frac{\gamma^{2}}{\left[v_{g}+v_{\rm 
Hxc}(n_{d})\right]^{2}+\gamma^{2}},
\label{ksg}
\ee
where $G_{0}=1/\pi$ is 
the quantum of conductance.
The exact KS conductance equals the exact conductance due to the Friedel sum 
rule and the WBLA.\cite{mera}
It can easily be seen that for $v_{\rm Hxc}=v^{\theta}_{\rm Hxc}$ the 
conductance is
correctly peaked at $v_{g}=\bar{v}_{g}$ but its width is weakly 
dependent on $U$. Indeed $v_{\rm Hxc}^{\theta}(n_{d})=U/2$ everywhere 
except that at the occupations $n_{d}=0,2$, see Fig. \ref{vxc}. Therefore the conductance 
as a function of $v_{g}$ has a constant width $\gamma$ since 
$n_{d}$ is never exactly 0 or 2. This is illustrated in Fig. 
\ref{comparison} where the conductance calculated using  $v_{\rm Hxc}^{\theta}$ is compared 
with the variational results of Ref. \onlinecite{mravlje}. 

\begin{figure}[tbp]
\includegraphics[width=6.2cm]{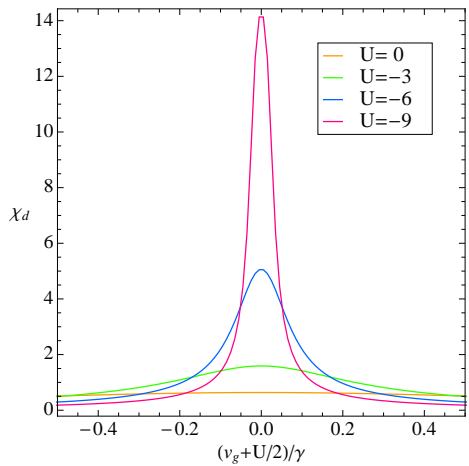}
\caption{Charge susceptibility at the impurity
$\chi_{d}=-\partial n_{d}/\partial v_{g}$
as a function of $v_{g}$ with potential $v^{l}_{\rm Hxc}$
for different values of $U$ (in units of $\gamma$).}
\label{compres}
\end{figure}

The potential $v_{\rm Hxc}^{\theta}$ can be substantially improved by following the observation of Ref. 
\onlinecite{evers}. At temperatures $T=1/\b$ below the Kondo temperature the 
broadening of the discontinuity in 
$v_{\rm Hxc}^{\theta}$ is proportional to $\b U e^{-\b U/2}$ and approaches zero 
in the limit $T\ra 0$. However, the exact 
Hartree-XC potential should have
an {\it intrinsic} broadening $W\sim 8 \g /\pi |U|$ due
to the finite hybridization of the impurity.
Therefore we here propose a Hartree-XC potential which is the convolution of 
$v^{\theta}_{\rm Hxc}$ with a Lorenzian of width $W$.
The resulting potential for negative $U$ and zero temperature reads
\be
v_{\rm Hxc}^{l}(n_{d})=\frac{U}{2}+\frac{U}{\pi}(\arctan 
\frac{n_{d}-2}{W}+\arctan \frac{n_{d}}{W}) .
\ee
In Fig. \ref{vxc} we show the comparison between $v^{\theta}_{\rm 
Hxc}$ at finite temperature and $v^{l}_{\rm Hxc}$ at zero 
temperature. Choosing $\b=1/W\gamma=\pi |U|/(8 \g^{2})$ we see that the thermal 
broadening  is much smaller than the Lorentzian 
broadening for $\g\ll |U|$. 
Figure \ref{comparison} clearly illustrates the crucial role of the 
broadening of the discontinuity in  the 
 shrinkage of the conductance resonance as $|U|$ increases. The figure  
displays also the conductance in the Hartree-Fock (HF) approximation, i.e., 
with potential $v_{\rm HF}(n_{d})=Un_{d}/2$.
Even though this potential  reproduces the shrinkage up to $U\sim 
-2$, it becomes unreliable already for $U\sim -3$. At this 
critical value the self-consistent equation for the density develops 
multiple solutions, three in our case, as shown in the bottom panel 
of Fig. \ref{comparison}. 
This multistability scenario should be contrasted with 
the positive-$U$ 
Anderson model where multiple solutions within the Hartree-Fock approximation 
are found only out-of-equilibrium.\cite{bistab}

Finally we used the Hartree-XC potential $v^{l}_{\rm Hxc}$ to calculate 
the charge susceptibility $\chi_{d}=-\partial n_{d}/\partial v_{g}$.  In Fig. 
\ref{compres} we show  $\chi_{d}$ as a function of $v_{g}$
for several values of $U$. Also in this case our approximation correctly 
captures the growth of $\chi_{d}$ at $v_{g}=\bar{v}_{g}$, another
typical signature of the charge-Kondo effect. We further observe 
that the 
hight of the peak in $\chi_{d}$ saturates to values around 2 if 
we use the Hartree-XC potential $v^{\theta}_{\rm Hxc}$ (not shown). 
Thus the broadening of the discontinutity is crucial in this case as 
well.

In conclusion we generalized the variational energy functional of  
ensemble-DFT to cases where the convexity inequality is not fulfilled.
The energy $E(N)$ is a convex function 
of the fractional particle number $N$, and it is given by  
the lowest series of straight lines joining a subset of ground-state 
energies. We discussed the 
relevance of this property in the description of correlated systems with attractive 
interactions. As for odd $N$ the energy $E(N)$ has no cusp and the KS 
discontinuity is zero, the XC discontinuity  
is  zero in these cases. We showed that  
the missing XC discontinuity and the broadening induced by the 
finite hybridization with the leads are essential features of any approximate 
functional to describe the 
charge-Kondo effect in the negative-$U$ Anderson model within 
ensemble-DFT. The functional proposed in this work yields results in 
fairly good agreement with
the available numerical data.
In particular the shrinkage of the conductance peak as well as 
the growth of the charge susceptibility with increasing $|U|$ 
are correctly captured.

\end{document}